\documentclass[longauth]{aa} 

\usepackage{graphicx}
\usepackage{txfonts}
\usepackage{graphicx}
\usepackage{nth}
\usepackage{longtable}
\usepackage{color}
\usepackage[draft]{hyperref}
\usepackage{txfonts}
\usepackage{natbib}
\bibpunct{(}{)}{,}{a}{}{,}

\usepackage[caption=false]{subfig}

\newcommand{\Msolar}{\mbox{${M}_{\sun}$}}
\newcommand{\Rsolar}{\mbox{${R}_{\sun}$}}

\newcommand{\Mjup}{\mbox{${M}_{J}$}}

\newcommand{\Rjup}{\mbox{${R}_{J}$}}
\newcommand{\rhojup}{\mbox{$\rho_{J}$}}

\newcommand{\teff}{\mbox{$T_{\rm eff}$}}
\newcommand{\logg}{\mbox{$\log g$}}

\newcommand\T{\rule{0pt}{2.2ex}}

\begin{document}

   \title{The hot dayside and asymmetric transit of WASP-189\,b seen by CHEOPS\thanks{The 
photometric time series data are only available in electronic form
at the CDS via anonymous ftp to cdsarc.u-strasbg.fr (130.79.128.5) or via 
http://cdsweb.u-strasbg.fr/cgi-bin/qcat?J/A+A/}}


\author{M.~Lendl\inst{\ref{inst1},\ref{inst2}}\relax
\and Sz.~Csizmadia\inst{\ref{inst14}}\relax
\and A.~Deline\inst{\ref{inst1}}\relax
\and L.~Fossati\inst{\ref{inst2}}\relax
\and D.~Kitzmann\inst{\ref{inst32}}\relax
\and K.~Heng\inst{\ref{inst32}}\relax
\and S.~Hoyer\inst{\ref{inst4}}\relax
\and S.~Salmon\inst{\ref{inst1},\ref{inst5}}\relax
\and W.~Benz\inst{\ref{inst3},\ref{inst32}}\relax
\and C.~Broeg\inst{\ref{inst3}}\relax
\and D.~Ehrenreich\inst{\ref{inst1}}\relax
\and A.~Fortier\inst{\ref{inst3}}\relax
\and D.~Queloz\inst{\ref{inst1},\ref{inst6}} \relax
\and A.~Bonfanti\inst{\ref{inst2}} \relax
\and A.~Brandeker\inst{\ref{inst10}}\relax
\and A.~Collier~Cameron\inst{\ref{inst9}} \relax 
\and L.~Delrez\inst{\ref{inst5},\ref{inst39},\ref{inst1}}\relax
\and A.~Garcia~Mu\~noz\inst{\ref{inst7}}\relax
\and M.J.~Hooton\inst{\ref{inst3}}\relax
\and P.F.L.~Maxted\inst{\ref{inst8}}\relax
\and B.M.~Morris\inst{\ref{inst32}}\relax
\and V.~Van~Grootel\inst{\ref{inst5}}\relax
\and T.G.~Wilson\inst{\ref{inst9}}\relax
\and Y.~Alibert\inst{\ref{inst3}}\relax
\and R.~Alonso\inst{\ref{inst11},\ref{inst37}}\relax
\and J.~Asquier\inst{\ref{inst31}}\relax
\and T.~Bandy\inst{\ref{inst3}}\relax
\and T.~B\'arczy\inst{\ref{inst24}}\relax
\and D.~Barrado\inst{\ref{inst29}}\relax
\and S.C.C~Barros\inst{\ref{inst16},\ref{inst38}}\relax
\and W.~Baumjohann\inst{\ref{inst2}}\relax
\and M.~Beck\inst{\ref{inst1}}\relax
\and T.~Beck\inst{\ref{inst32}}\relax
\and A.~Bekkelien\inst{\ref{inst1}}\relax
\and M.~Bergomi\inst{\ref{inst21}}\relax
\and N.~Billot\inst{\ref{inst1}}\relax
\and F.~Biondi\inst{\ref{inst21}}\relax
\and X.~Bonfils\inst{\ref{inst13}}\relax
\and V.~Bourrier\inst{\ref{inst1}}\relax
\and M-D.~Busch\inst{\ref{inst3}}\relax
\and J.~Cabrera\inst{\ref{inst14}}\relax
\and V.~Cessa\inst{\ref{inst32}}\relax
\and S.~Charnoz\inst{\ref{inst15}}\relax
\and B.~Chazelas\inst{\ref{inst1}}\relax
\and C.~Corral Van Damme\inst{\ref{inst31}}\relax
\and M.B.~Davies\inst{\ref{inst26}}\relax
\and M.~Deleuil\inst{\ref{inst4}}\relax
\and O.D.S~Demangeon\inst{\ref{inst4},\ref{inst16}}\relax
\and B.-O.~Demory\inst{\ref{inst32}}\relax
\and A.~Erikson\inst{\ref{inst14}}\relax
\and J.~Farinato\inst{\ref{inst21}}\relax
\and M.~Fridlund\inst{\ref{inst17},\ref{inst36}}\relax
\and D.~Futyan\inst{\ref{inst1}}\relax
\and D.~Gandolfi\inst{\ref{inst18}}\relax
\and M.~Gillon\inst{\ref{inst39}}\relax
\and P.~Guterman\inst{\ref{inst4},\ref{inst33}}\relax
\and J.~Hasiba\inst{\ref{inst2}}\relax
\and E.~Hernandez\inst{\ref{inst32}}\relax
\and K.G.~Isaak\inst{\ref{inst31}}\relax
\and L.~Kiss\inst{\ref{inst23}}\relax
\and T.~Kuntzer\inst{\ref{inst1}}\relax
\and A.~Lecavelier~des~Etangs\inst{\ref{inst25}}\relax
\and T.~L\"uftinger\inst{\ref{inst19}}\relax
\and J.~Laskar\inst{\ref{inst20}}\relax
\and C.~Lovis\inst{\ref{inst1}}\relax
\and D.~Magrin\inst{\ref{inst21}}\relax
\and L.~Malvasio\inst{\ref{inst32}}\relax
\and L.~Marafatto\inst{\ref{inst21}}\relax
\and H.~Michaelis\inst{\ref{inst14}}\relax
\and M.~Munari\inst{\ref{inst22}}\relax
\and V.~Nascimbeni\inst{\ref{inst21}}\relax
\and G.~Olofsson\inst{\ref{inst10}}\relax
\and H.~Ottacher\inst{\ref{inst2}}\relax
\and R.~Ottensamer\inst{\ref{inst19}}\relax
\and I.~Pagano\inst{\ref{inst22}}\relax
\and E.~Pall\'e\inst{\ref{inst11},\ref{inst37}}\relax
\and G.~Peter\inst{\ref{inst34}}\relax
\and D.~Piazza\inst{\ref{inst3}}\relax
\and G.~Piotto\inst{\ref{inst42},\ref{inst21}}\relax
\and D.~Pollacco\inst{\ref{inst27}}\relax
\and F.~Ratti\inst{\ref{inst31}}\relax
\and H.~Rauer\inst{\ref{inst14},\ref{inst7},\ref{inst41}}\relax
\and R.~Ragazzoni\inst{\ref{inst21}}\relax
\and N.~Rando\inst{\ref{inst31}}\relax
\and I.~Ribas\inst{\ref{inst12},\ref{inst35}}\relax
\and M.~Rieder\inst{\ref{inst3}}\relax
\and R.~Rohlfs\inst{\ref{inst1}}\relax
\and F.~Safa\inst{\ref{inst31}}\relax
\and N.C.~Santos\inst{\ref{inst16},\ref{inst38}}\relax
\and G.~Scandariato\inst{\ref{inst22}}\relax
\and D.~S\'egransan\inst{\ref{inst1}}\relax
\and A.E.~Simon\inst{\ref{inst3}}\relax
\and V.~Singh\inst{\ref{inst22}}\relax
\and A.M.S.~Smith\inst{\ref{inst14}}\relax
\and M.~Sordet\inst{\ref{inst1}}\relax
\and S.G.~Sousa\inst{\ref{inst16}}\relax
\and M.~Steller\inst{\ref{inst2}}\relax
\and Gy.M.~Szab\'o\inst{\ref{inst30},\ref{inst40}}\relax
\and N.~Thomas\inst{\ref{inst3}}\relax
\and M.~Tschentscher\inst{\ref{inst14}}\relax
\and S.~Udry\inst{\ref{inst1}}\relax
\and V.~Viotto\inst{\ref{inst21}}\relax
\and I.~Walter\inst{\ref{inst34}}\relax
\and N.A.~Walton\inst{\ref{inst28}}\relax
\and F.~Wildi\inst{\ref{inst1}}\relax
\and D.~Wolter\inst{\ref{inst14}}\relax
}

\institute{Observatoire de Gen\`eve, Universit\'e de Gen\`eve, Chemin des maillettes 51, 1290 Sauverny, Switzerland; \email{monika.lendl@unige.ch}\label{inst1}\relax
\and Space Research Institute, Austrian Academy of Sciences, Schmiedlstr. 6, 8042 Graz, Austria \label{inst2}\relax
\and Institute of Planetary Research, German Aerospace Center (DLR), Rutherfordstr. 2, 12489, Berlin, Germany \label{inst14}\relax
\and Center for Space and Habitability, Gesellsschaftstr. 6, 3012, Bern, Switzerland \label{inst32}\relax
\and Laboratoire d’Astrophysique de Marseille, Univ. de Provence, UMR6110 CNRS, 38 r. F. Joliot Curie, 13388 Marseille, France \label{inst4}\relax
\and Physikalisches Institut, University of Bern, Gesellschaftsstr. 6, 3012 Bern, Switzerland \label{inst3}\relax
\and Space sciences, Technologies and Astrophysics Research (STAR) Institute, Universit{\'e} de Li{\`e}ge, All{\'e}e du 6 Ao{\^u}t 17, 4000 Li{\`e}ge, Belgium \label{inst5}\relax
\and Astrophysics Group, Cavendish Laboratory, J.J. Thomson Avenue, Cambridge CB3 0HE, United Kingdom \label{inst6}\relax
\and Astrophysics Group, Keele University, Staffordshire, ST5 5BG, United Kingdom \label{inst8}\relax
\and Department of Astronomy, Stockholm University, AlbaNova University Center, 10691 Stockholm, Sweden \label{inst10}\relax
\and School of Physics and Astronomy, Physical Science Building, North Haugh, St Andrews, United Kingdom \label{inst9}\relax
\and Astrobiology Research Unit, Université de Liège, Allée du 6 Ao\^ut 19C, B-4000 Liège, Belgium \label{inst39}\relax
\and Center for Astronomy and Astrophysics, Technical University Berlin, Hardenbergstr. 36, 10623 Berlin, Germany \label{inst7}\relax
\and Instituto de Astrof\'isica de Canarias (IAC), 38200 La Laguna, Tenerife, Spain \label{inst11}\relax
\and Deptartamento de Astrof\'isica, Universidad de La Laguna (ULL), E-38206 La Laguna, Tenerife, Spain \label{inst37}\relax
\and ESTEC, European Space Agency, Keplerlaan 1, 2201 AZ Noordwijk, The Netherlands \label{inst31}\relax
\and Admatis, Miskolc, Hungary \label{inst24}\relax
\and Depto. de Astrof\'{\i}sica, Centro de Astrobiolog\'{\i}a (CSIC-INTA), ESAC campus, 28692 Villanueva de la C\~ada (Madrid), Spain \label{inst29}\relax
\and Instituto de Astrof\'isica e Ci\^encias do Espa\c{c}o, Universidade do Porto, CAUP, Rua das Estrelas, 4150-762 Porto, Portugal \label{inst16}\relax
\and Departamento de F\'isica e Astronomia, Faculdade de Ci\^encias, Universidade do Porto, Rua do Campo Alegre, 4169-007 Porto, Portugal \label{inst38}\relax
\and INAF, Osservatorio Astronomico di Padova, Vicolo dell’Osservatorio 5, 35122, Padova, Italy \label{inst21}\relax
\and Universit\'e Grenoble Alpes, CNRS, IPAG, 38000 Grenoble, France \label{inst13}\relax
\and Institut de Physique du Globe de Paris (IPGP), 1 rue Jussieu, 75005, Paris, France \label{inst15}\relax
\and Lund Observatory, Dept. of Astronomy \& Theoretical Physics, Lund University, Box 43, Lund, 22100, Sweden \label{inst26}\relax
\and Leiden Observatory, University of Leiden, PO Box 9513, 2300 RA, Leiden, The Netherlands \label{inst17}\relax
\and Department of Space, Earth and Environment, Chalmers University of Technology, Onsala Space Observatory, 439 92 Onsala, Sweden \label{inst36}\relax
\and INAF, Osservatorio Astrofisico di Torino, via Osservatorio 20, 10025, Pino Torinese, Italy \label{inst18}\relax
\and Division Technique INSU, BP 330, 83507 La Seyne cedex, France \label{inst33}\relax
\and Konkoly Observatory, Research Centre for Astronomy and Earth Sciences, 1121 Budapest, Konkoly Thege Mikl\'os \'ut 15-17, Hungary \label{inst23}\relax
\and Institut d’astrophysique de Paris, UMR7095 CNRS, Universit\'e Pierre \& Marie Curie, 98bis blvd. Arago, 75014 Paris, France \label{inst25}\relax
\and University of Vienna, Department of Astrophysics, T\"urkenschanzstr. 17, 1180 Vienna, Austria \label{inst19}\relax
\and IMCCE, UMR8028 CNRS, Observatoire de Paris, PSL Univ., Sorbonne Univ., 77 av. Denfert-Rochereau, 75014 Paris, France \label{inst20}\relax
\and INAF, Osservatorio Astrofisico di Catania, Via S. Sofia 78, 95123, Catania, Italy \label{inst22}\relax
\and Institute of Optical Sensor Systems, German Aerospace Center (DLR), Rutherfordstr. 2, 12489 Berlin, Germany \label{inst34}\relax
\and Dipartimento di FIsica e Astronomia "Galileo Galilei", Universita' degli Studi di Padova, Vicolo dell'Osservatorio 3, 35122 Padova, Italia \label{inst42}\relax
\and Department of Physics, University of Warwick, Gibbet Hill Road, Coventry CV4 7AL, United Kingdom \label{inst27}\relax
\and Institut f\"ur Geologische Wissenschaften, Freie Universit\"at Berlin, 12249 Berlin, Germany \label{inst41}\relax
\and Institut de Ci\`encies de l’Espai (ICE, CSIC), Campus UAB, C/CanMagrans s/n, 08193 Bellaterra, Spain \label{inst12}\relax
\and Institut d'Estudis Espacials de Catalunya (IEEC), Gran Capit\`a 2-4, 08034 Barcelona, Spain \label{inst35}\relax
\and ELTE E\"otv\"os Lor\'and University, Gothard Astrophysical Observatory, Szombathely, Hungary \label{inst30}\relax
\and MTA-ELTE Exoplanet Research Group, 9700 Szombathely, Szent Imre h. u. 112, Hungary \label{inst40}\relax
\and Institute of Astronomy, University of Cambridge, Madingley Road, Cambridge CB3 0HA, United Kingdom \label{inst28}\relax
}    
   
   \date{}

  \abstract
  {
  The CHEOPS space mission dedicated to exoplanet follow-up was launched in December 2019, equipped with the capacity to perform photometric 
  measurements at the 20\,ppm level. As CHEOPS carries out its observations in a broad optical passband, it can provide insights 
  into the reflected light from exoplanets and constrain the short-wavelength thermal emission for the hottest of planets 
  by observing occultations and phase curves.
  
  Here, we report the first CHEOPS observation of an occultation, namely, that of the hot Jupiter WASP-189\,b, a 
  $M_P \approx 2 M_J$ planet orbiting an A-type star.  
  We detected the occultation of WASP-189\,b at high significance in 
  individual measurements and derived an occultation depth of $dF = 87.9 \pm 4.3$~ppm based on four occultations.
  We compared these measurements to model predictions and we find that 
  they are consistent with an unreflective atmosphere heated to a temperature of $3435 \pm 27$\,K, when
assuming inefficient heat redistribution. 
  
  Furthermore, we present two transits of WASP-189\,b observed by CHEOPS. These transits have an asymmetric shape that we
  attribute to gravity darkening of the host star caused by its high rotation rate. We used these measurements to refine the planetary 
  parameters, finding a $\sim25\%$ deeper transit compared to the discovery paper and updating the radius of WASP-189\,b to $1.619\pm0.021 R_J$. 
  We further measured the projected orbital obliquity to be $\lambda = 86.4^{+2.9}_{-4.4}$\,deg, a value that is in good agreement with 
  a previous measurement from spectroscopic observations, and derived a true obliquity of $\Psi = 85.4\pm4.3$\,deg.
  
  Finally, we provide reference values for the photometric precision attained by the CHEOPS satellite: for the V=6.6 mag star, and using a one-hour binning, 
  we obtain a residual RMS between 10 and 17~ppm on the individual light curves, and 5.7~ppm when combining the four visits.
  }


   \maketitle
%

\section{Introduction}
\label{sec:intro}
The CHaracterising ExOPlanets Satellite (CHEOPS) is the first European space mission dedicated primarily to the study of known
extrasolar planets. It consists of a 30~cm (effective) aperture telescope collecting ultra-high precision time-series photometry of 
exoplanetary systems in a broad optical passband \citep{Benz20}. 
Unlike the previous space observatories dedicated to exoplanets, CoRoT \citep{Baglin06}, Kepler \citep{Borucki10}, K2 \citep{Howell2014},
and the ongoing  TESS mission \citep{Ricker14}, CHEOPS is a pointed mission, optimised to obtain high-cadence photometric observations
at the 20~ppm level for a single star at a time.
CHEOPS was launched successfully into a 700~km altitude Sun-synchronous polar orbit on 18 December 2019 
and its first science observations were obtained in late March 2020. 

As one of its first scientific targets, CHEOPS observed the ultra-hot Jupiter WASP-189\,b \citep{Anderson18}, a gas giant transiting the 
bright ($V=6.6$\,mag) A-type star HD\,133112. WASP-189\,b is one of the most highly irradiated planets known thus far, with a 
dayside equilibrium temperature of $\sim3400$\,K \citep{Anderson18}.
It orbits an early-type star similarly to the extreme object KELT-9b \citep{Gaudi17}, 
but with a longer orbital period of 2.7\,days, placing it closer, in temperature, to ultra-short period planets orbiting F and G stars. As such, this object allows 
us to comparatively probe the impact of different stellar spectral energy distributions and, in particular, strong short-wavelength irradiation on planetary atmospheres. 
As it is orbiting around an A-type star, the system  is also relatively young ($730\pm130$~Myr, see Section \ref{sec:star}), providing us with a window into the atmospheric evolution of close-in gas giants.

In this paper, we report on CHEOPS observations of four occultations and two transits of WASP-189\,b. We use the occultations to constrain the planet's
temperature and reflective properties and the transits to revise the planetary radius and determine the system's orbital obliquity from the gravity darkening of the host star
and the associated light curve asymmetry.
We describe the observations and data reduction in Section \ref{sec:obs}, discuss the results in Section \ref{sec:res}, and present a brief conclusion in Section \ref{sec:con}.

\section{Observations, data reduction, and analysis}
\label{sec:obs}

\subsection{CHEOPS observations of WASP-189\,b}
\label{sec:obsCH}

We observed four occultations of WASP-189\,b between 19 March and 7 April 2020. The individual observations lasted between 12.4 and 13 h, 
distributed over either seven or eight spacecraft orbits of 98.77\,min, thus covering the 3.35\,h occultation, together with substantial out-of-eclipse baseline. 
During the analysis of the occultation data, we obtained further observations of two transits of WASP-189\,b with CHEOPS on 15 and 18 June 2020, 
which we subsequently included in the final analysis. The transit observations covered the transit, together with a total of six CHEOPS orbits obtained outside of it. 
The observations were interrupted for up to 41 and up to 17 min per orbit due to Earth occultations or passages through the 
South Atlantic Anomaly (SAA), respectively. These instances can be seen as gaps in the light curves displayed in Figures \ref{fig:lcs} and \ref{fig:lctrans}. We used exposure times of 4.8~s and co-added, on board, seven individual exposures of the G=6.55 mag star,
resulting in an effective cadence of 33.4\,s. 
A full description of the CHEOPS telescope and the technical details of its observations is presented in \citet{Benz20}.

The data were processed with the CHEOPS data reduction pipeline \citep[DRP,][]{Hoyer20}, which performs image correction and uses aperture photometry 
to extract target fluxes for various apertures. 
The CHEOPS DRP was thoroughly tested, both using the CHEOPS data simulator \citep{Futyan20} and data obtained during commissioning. Using simulated data,
we performed a series of injection and retrieval tests covering a range of planetary transit scenarios and levels of field crowding. The data obtained during the
commissioning consisted of observations of stable stars that confirmed the stability of the photometry in the presence of interruptions due to SAA crossings and Earth occultations.
During commissioning, we also carried out transit observations and verified that the retrieved transit parameters were in good agreement with literature values \citep[see e.g.][]{Benz20}.
For the occultations and the transits, versions 11 and 12 of the DRP were used, respectively. We found a minimal light curve 
RMS for the default aperture of 25~pixels.

\begin{figure}
                \includegraphics[width=\linewidth]{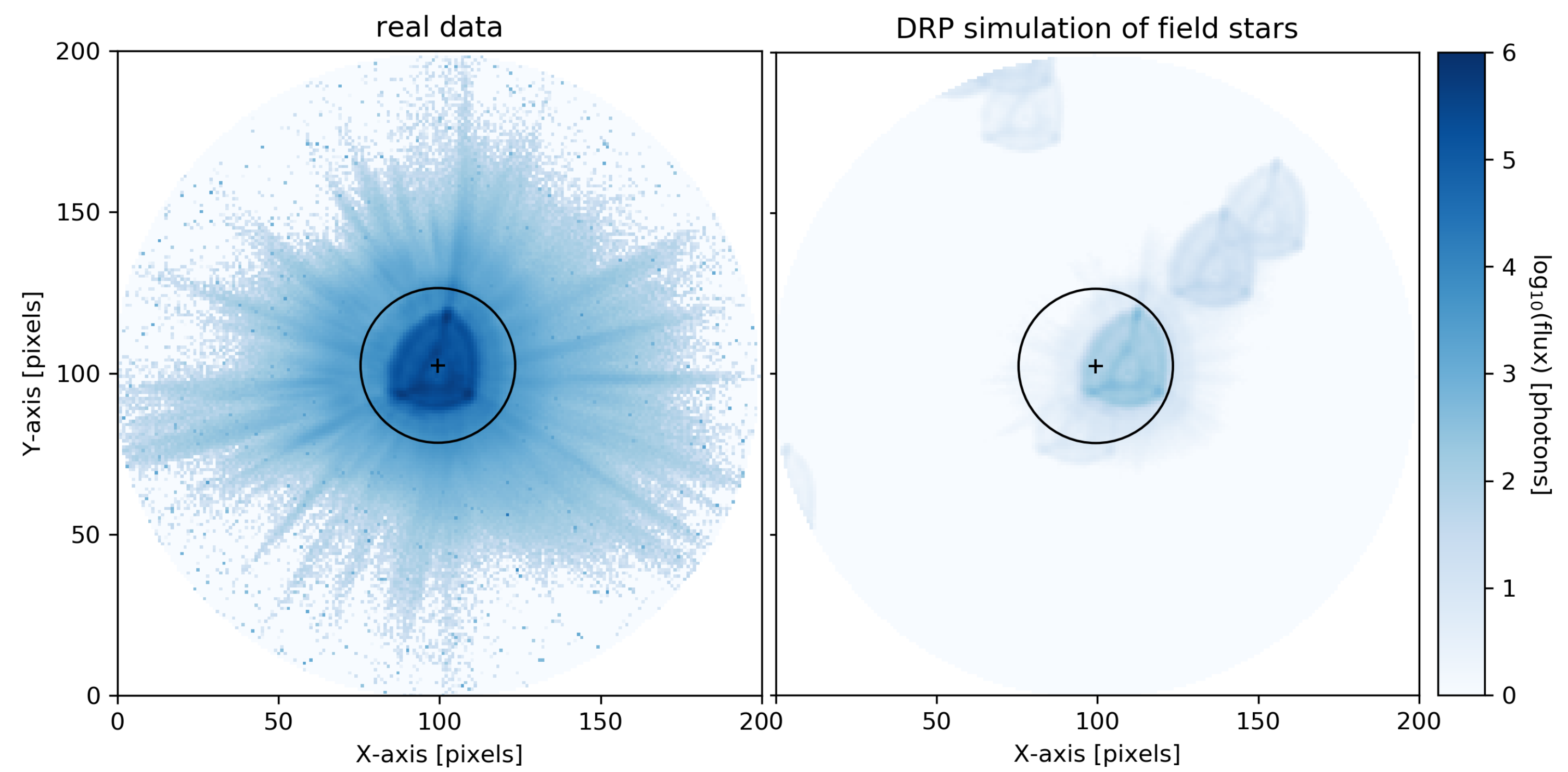}
                \caption{Example of the field of view of WASP-189 observed by CHEOPS (left) and its respective DRP simulation with the target removed (right). 
                The circle and the cross represent the photometric aperture and the location of the target's PSF, respectively. The triangular shape of the CHEOPS PSF is clearly visible. }
                \label{fig:fig_fov_data_sims}
\end{figure}

Owing to the extended and irregular shape of CHEOPS' point spread function (PSF) and the fact that the field rotates around the target along the satellite's orbit,
nearby stars produce a time-variable flux contamination in the photometric aperture, in phase with the spacecraft's roll angle. As explained in \cite{Hoyer20}, the DRP automatically 
determines the level of such contamination in the target's aperture for each exposure. The contamination is estimated from simulated 
images \citep{Futyan20} that are based on the CHEOPS PSF, the roll angle of each image. and the {\it Gaia} DR2 \citep{GaiaCollaboration2018} coordinates and magnitudes 
of all the stars with G<19.5\,mag in the field of view. In order to determine the level of contamination, our simulations were created both with and without the target.
Due to its brightness, WASP-189 appears to be well-isolated in the observed data, but the simulations show two faint contaminating sources 
located inside the aperture, with \emph{Gaia} G magnitudes of 14.4 and 18.9 and distances of 9 and 19 arcsec from the target, respectively. 
Figure \ref{fig:fig_fov_data_sims} shows a typical observation, as well as the corresponding simulated image containing only the background sources. 
We used these simulations to compute the time-variable contamination in the photometric aperture, finding that it is in excellent agreement with 
the observed flux variations on the CHEOPS orbital time scale. This allowed us to correct our photometric measurements for contamination (see Section \ref{sec:dat}).

\subsection{Host star properties}
\label{sec:star}
To assist in our analysis of the WASP-189 system, we derived fundamental stellar parameters via spectral line and spectral energy distribution (SED) fitting, along with stellar evolution modelling.
We estimated the stellar atmospheric parameters by comparing an average of 17 archival HARPS spectra with synthetic spectra 
computed using the {\it synth3} code \citep{kochukhov2007}, employing the tools described in \citet{fossati2007}.
We computed stellar atmosphere models using LLmodels \citep{shulyak2004}.
We used an iterative procedure to derive the effective temperature ($T_{\rm eff}$) by imposing excitation equilibrium for both 57 FeI and 
10 FeII lines, the surface gravity (log\,$g$) by imposing Fe ionisation equilibrium, and the microturbulence velocity ($\nu_{\rm mic}$) by minimising 
the standard deviation in the Fe abundance. Prior to fitting the lines, we measured the stellar projected rotational velocity 
($\nu$sin$I_\ast = $~93.1$\pm$1.7\,km\,s$^{-1}$) from several unblended lines.
We confirmed this measurement by applying the Fourier analysis technique \citep{gray2005,murphy2016}  to a handful of unblended lines. 
We find $T_{\rm eff}$\,=\,8000$\pm$80\,K, log\,$g$\,=\,3.9$\pm$0.2, and $\nu_{\rm mic}$\,=\,2.7$\pm$0.3\,km\,s$^{-1}$. Both $T_{\rm eff}$ and log\,$g$ are in good agreement with those derived by \citet{Anderson18}. 
We measured an iron abundance [Fe/H] of +0.29$\pm$0.13\,dex, as well as the abundances of C, O, Na, Mg, Si, S, Ca, Sc, Ti, Cr, Ni, Y, and Ba, obtaining the pattern shown 
in Appendix~\ref{app:abundances}.

The derived abundance pattern is typical of chemically peculiar metallic-line (Am) stars \citep{fossati2007,fossati2008},
which are limited to stars with a rotational velocity lower than $\approx$100\,km\,s$^{-1}$ \citep{michaud1970}. 
Therefore, as the measured stellar $\nu$sin$I_\ast$ value is close to the maximum rotational velocity for which Am chemical peculiarities can arise, 
the stellar inclination angle should be 
close to 90\,deg. 
The peculiar abundance pattern characterises only the stellar atmosphere and does not reflect the internal abundances, 
which we estimate at $+$0.2\,dex from the abundances of Mg, Si, and S -- elements that have been shown to be a good probe of the internal stellar metallicity \citep{fossati2007,fossati2008}.

In order to determine the stellar radius of WASP-189, we utilised the infrared flux method (IRFM; \citealt{Blackwell1977}), which permits the 
calculation of stellar angular diameter and $T_{\rm eff}$ using previously derived relations between these parameters and optical 
and infrared broadband fluxes as well as the synthetic photometry conducted on stellar atmospheric models over the bandpasses of the observed data.
We retrieved fluxes and corresponding uncertainties in the {\it Gaia} G, G$_{\rm BP}$, and G$_{\rm RP}$, 2MASS J, H, and K, and {\it WISE} W1 and W2 bandpasses 
taken from the most recent data release archives, respectively \citep{Skrutskie2006,Wright2010,GaiaCollaboration2018}. Stellar synthetic models \citep{Castelli2003} 
were fitted to the obtained broadband photometry in a Markov Chain Monte Carlo (MCMC) approach, with priors on the stellar parameters taken from the spectroscopic analysis detailed above. 
The derived stellar angular diameter was combined with the {\it Gaia} parallax to determine the stellar radius, $R_{*,\rm IRFM} = 2.362 \pm 0.030 R_{\odot}$. 
This value is in good agreement with the value reported in the discovery paper \citep{Anderson18}, with a precision, in fact, that is four times greater. 

Finally, we used $T_{\rm eff}$, metallicity (using 0.2$\pm$0.1\,dex, see above), and $R_{*,\rm IRFM}$ as inputs to obtain stellar mass and age through stellar evolution modeling. We merged the results from 
two independent approaches and stellar evolution codes: the Li\`ege code CLES with a Levenberg-Marquardt approach, as in \citet{Buldgen16}, and the PARSEC code with the approach described in \citet{bonfanti15,bonfanti16}. We varied the input physics in 
stellar models (particularly with regard to the importance of convective overshooting and mixing of elements induced by diffusion) 
and we checked the consistency between our two approaches, which was found to be excellent. 
We ultimately infer a mass of $M_* = 2.030 \pm 0.066 M_{\odot}$ and an age of 730 $\pm$ 130 Myr. The stellar parameters are listed in Table \ref{tab:par}.

\subsection{CHEOPS Data analysis}
\label{sec:dat}

We initially carried out an analysis that included only the occultations observed during the first weeks of scientific operations. 
However, later transit observations evidently showed an unexpectedly deep transit. 
We included these new data in our analysis, as a well-measured planetary radius is needed to properly interpret the occultation signal.

In addition to the astrophysical signals, the light curves contain the effect of variable contamination, which introduces a V-shaped flux 
variation in phase with the spacecraft roll angle (clearly visible in Figure \ref{fig:lcs}). Furthermore, several visits show trends with 
time, the origin of which could lie in $\delta$ Scuti or $\gamma$ Doradus-type stellar pulsations. 

\subsubsection{Occultation}
\label{sec:occ}

\begin{figure*}
\centering
\includegraphics[width=\linewidth]{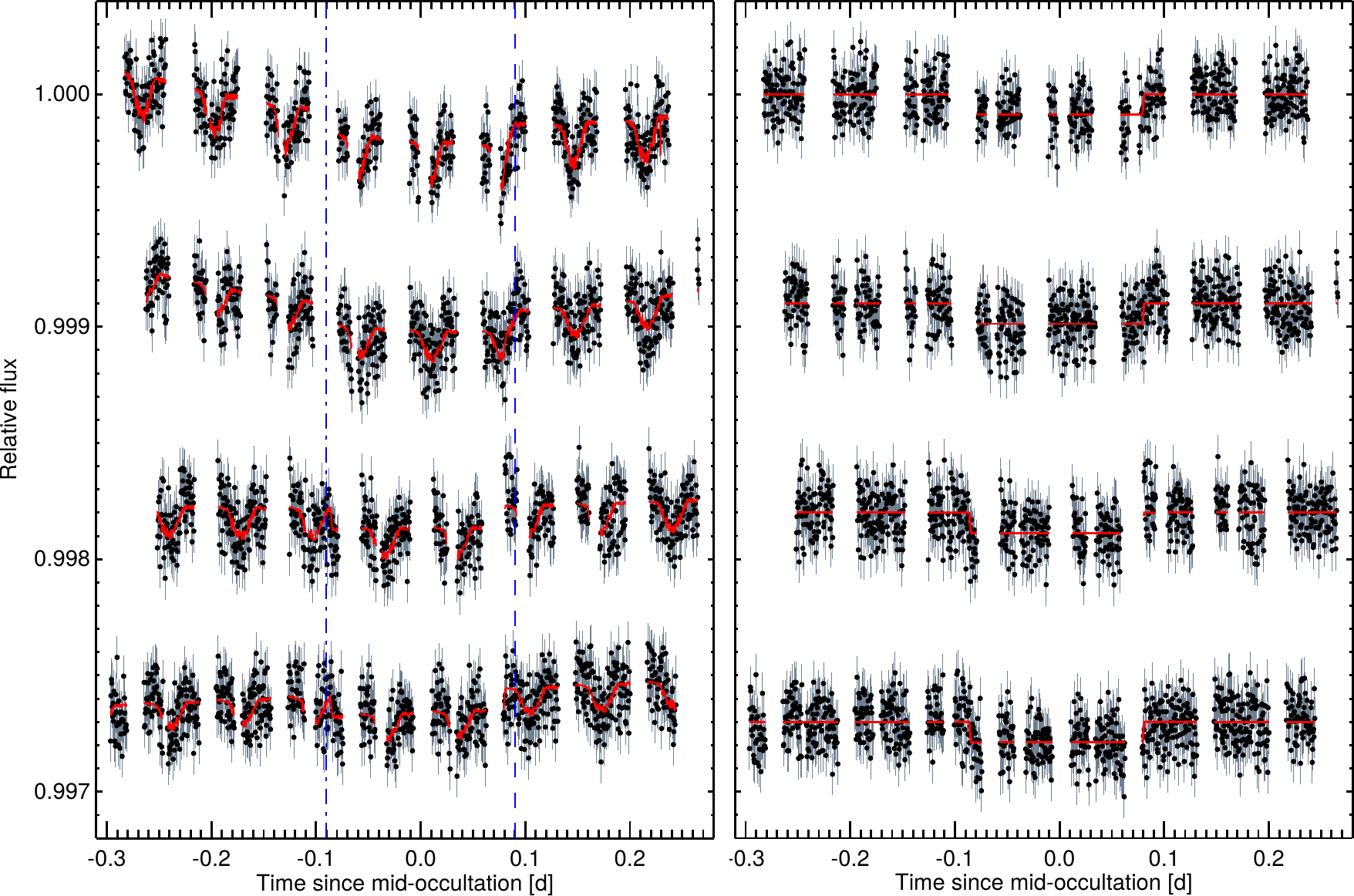}
\caption{Individual CHEOPS observations of four WASP-189\,b occultations. In both panels, 
visits are shown chronologically from top to bottom, occurring on 19, 27, and 30 March and 7 April 2020, respectively.
Left: Uncorrected observations (black points) together with their full (baseline and occultation, red line) light-curve models. 
Blue vertical dash-dotted lines indicate the beginning and end of occultation. Right: Data (black points) corrected for the 
instrumental and stellar trends, together with the occultation model (red line).}
\label{fig:lcs}
\end{figure*}

We carried out the analysis using an MCMC framework (\emph{CONAN}, \citealp{Lendl20a}), modeling the occultation signal at the same 
time as these signals of non-planetary origin to ensure a full propagation of uncertainties. To account for
correlated noise, we made use of either parametric models \citep[e.g.][]{Gillon10a} or Gaussian Processes \citep[GP; using the 
\emph{George} package][]{Ambikasaran14}, or a combination of both (i.e. using a parametric function multiplied with the transit model as the GP mean model).
To prescribe the occultation light curve, we used a limb-darkening-free \cite{Mandel02} transit model. 
To account for our knowledge of the planetary transit parameters, we placed Gaussian priors corresponding to the values and uncertainties found 
from the CHEOPS transits (see Section \ref{sec:transit}) on the impact parameter, $b,$ and the transit duration, $T_{\mathit{14}}$, the radius 
ratio, $R_P/R_\ast$. Uniform priors were assumed for the occultation depth, $dF_\mathit{occ}$, and the mid-transit time, $T_0$.
The period was kept fixed and the eccentricity was assumed
to be zero \citep[as found by][]{Anderson18}. For the radial velocity amplitude, $K$, and the stellar mass and radius ($M_\ast$, $R_\ast$), which 
are unconstrained by our analysis, we assumed Gaussian distributions, centred on the values of \citet{Anderson18} or, where appropriate, 
the values reported in Section \ref{sec:star}. 

\begin{figure*}
\centering
\includegraphics[width=\linewidth]{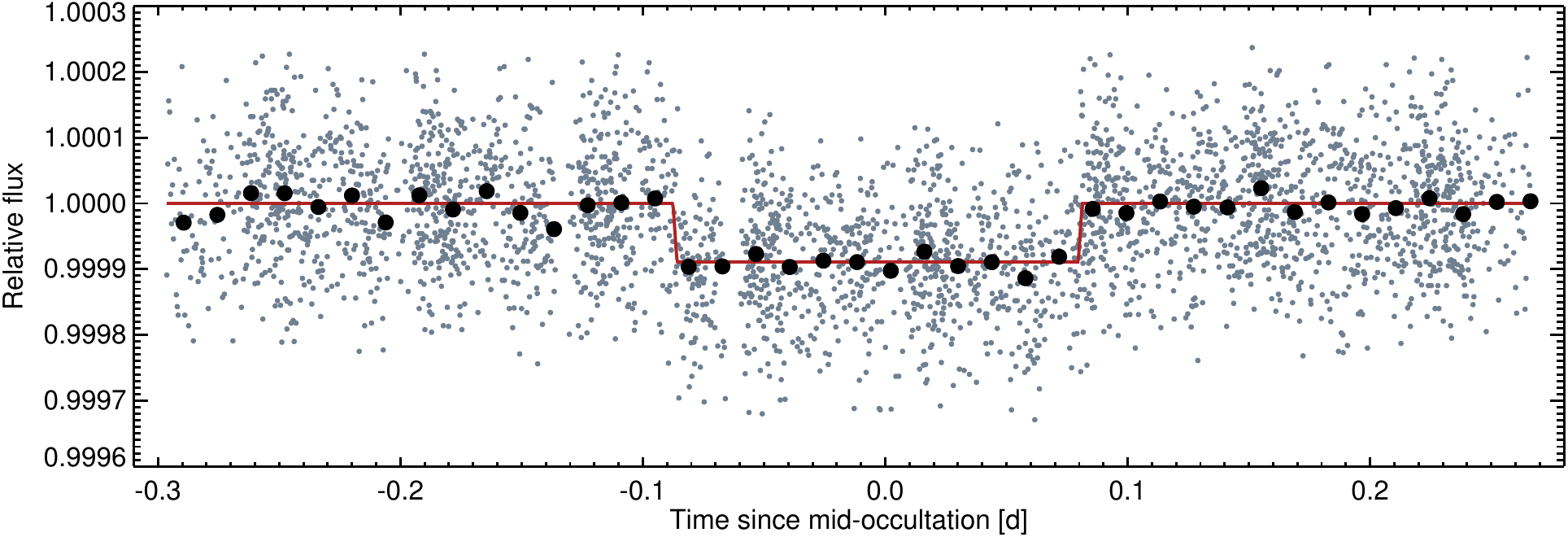}
\caption{Corrected and phase-folded CHEOPS occultation light curve of WASP-189\,b. Black points show the light curve binned into 20-minute intervals and the red line 
shows the final occultation model.}
\label{fig:clc}
\end{figure*}

We explored a large range of models for the correlated noise, testing both parametric models composed of polynomials up to \nth{4} order
in the recorded state variables (most importantly: time, PSF center, contamination, and spacecraft roll angle) as well as GPs using time, roll angle, 
and contamination, or a combination of these, as input. We tested both a \emph{Mat\'ern-3/2} and an \emph{exponential-squared} kernel. We find that the 
systematics are equally well-modeled by using either a combination of time polynomials (modeling the slow trends) 
paired with a \emph{Mat\'ern-3/2} GP with the telescope roll angle as input (modeling the contamination), or a combination of first- and second-order time
polynomials together with a linear dependence on the contamination value.
Both the results and derived uncertainties associated with each approach are fully compatible.
We selected the latter as our preferred model, as it accounts for our physical understanding of the source 
of the roll-angle-dependent variability. We report the results of our analysis in Table \ref{tab:par}. Individual light curves
are shown in Figure \ref{fig:lcs}, with the corrected and phase-folded data presented in Figure \ref{fig:clc}.

We also carried out an independent analysis using the {\tt pycheops}\footnote{\url{https://github.com/pmaxted/pycheops}} package, which
is being developed specifically for the analysis of CHEOPS data. Optimisation of the model parameters was done using 
{\tt lmfit}\footnote{\url{https://lmfit.github.io/lmfit-py/}} and detrending done either via a parametric method 
of decorrelating the data linearly against the contamination or roll angle, and quadratically against time, or 
 a GP regression with a \emph{Mat\'ern-3/2} kernel to model the flux against roll angle trend using the celerite package \citep{Foreman-Mackey17}.
Again, we obtained values that are fully compatible with the reported ones.

\subsubsection{Transit}
\label{sec:transit}

\begin{figure}
\centering
\includegraphics[width=9cm]{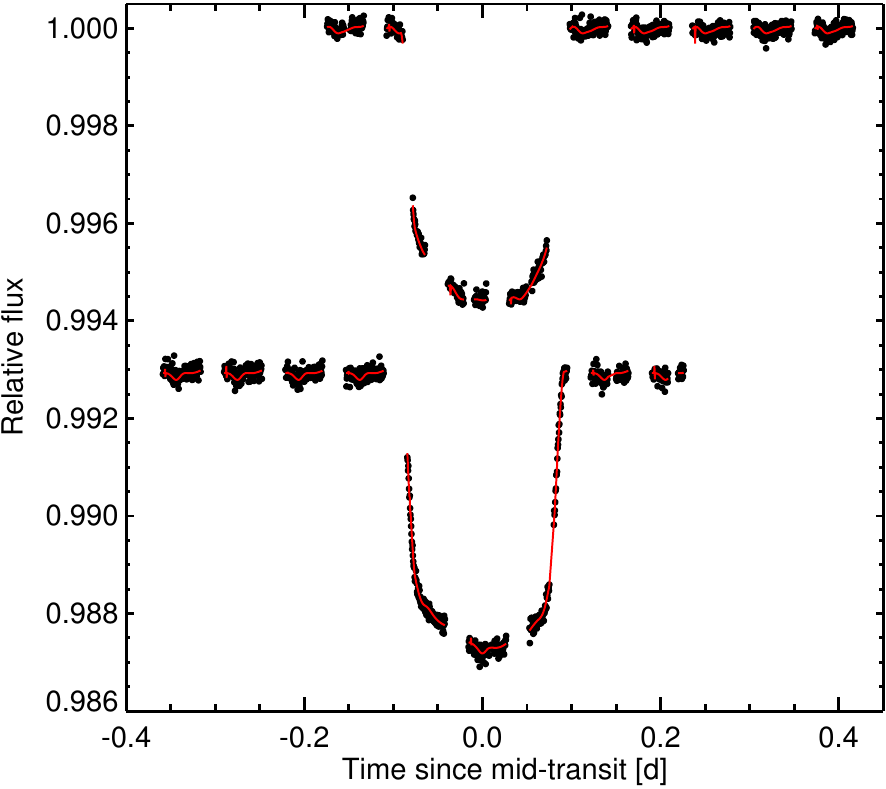}
\caption{ \label{fig:lctrans}Uncorrected CHEOPS observations of two transits of WASP-189\,b (black points), together with their full (baseline and transit) 
light-curve models (red lines). The upper light curve was observed on 15 June 2020 and the bottom light curve on 18 June 2020.}
\end{figure}

At the photometric precision reached by CHEOPS, the planetary transit can be seen to be asymmetric, a feature most 
readily explained by the presence of gravity darkening due to the combination of the host star's 
fast rotation and the planet's inclined orbit \citep{VonZeipel24,Barnes09}. 
Accounting for gravity darkening in transit models is computationally intensive and, therefore, we performed an independent analysis of the transits and used the results as priors for the analysis of the occultations (see Section \ref{sec:occ}).
We used the Transit and Light Curve Modeller (TLCM, see \citealp{Csizmadia20} for details) for this purpose. This code uses the analytic expressions of \citet{Mandel02} for 
the transit model and allows us to jointly model the transit together with various baseline models that account for correlated noise. 

To model the gravity darkening, we compute a modification to the analytic model taking into account the varying stellar flux emitted along the planet's transit path. To do so, the stellar surface is divided into 120x120 surface elements (in longitude and in latitude) and, for each, the surface effective temperature is calculated via
\begin{equation}
    T_{local} = T_{\ast} \left(\frac{|\nabla V|_{local}}{|\nabla V|_{pole}} \right)^{0.25}
.\end{equation}
We assume\ a polar temperature of $T_{pole} = 8000$ K and the above equation inherently assumes a gravity darkening exponent of 1.0, which is appropriate for hot stars \citet{Claret14}. The local surface gravitational potential ($V$) is calculated by assuming a two-axial ellipsoidal shape of the host star and given as\footnote{Stellar gravitational potential $V = GM/R_{\ast}$ was expressed by more easily measurable quantities via Kepler's third law.}  
\begin{equation}
    V = \frac{n^2 a^3}{(1+q) r} + \frac{1}{2} \omega_{rot}^2 r^2 \sin^2 b \, ,
\end{equation}
with the mass ratio, $q = M_{p} / M_{\ast}$, the mean motion, $n,$ and the astrographic latitude, $b$. The rotational angular velocity 
($\omega_{rot}$) is calculated from the stellar radius, $R_\star = 2.36 \pm 0.030$, the $\nu\sin I_\ast = 93.1\pm 1.7\,kms^{-1}$ (see Section \ref{sec:star}), 
and the fitted stellar inclination. 
We fit two angles: the inclination of the stellar rotational vector, $I_\ast$, and its tilt-angle relative to 
celestial north direction ($\Omega_{star} = 90^\circ - \lambda$). These two angles fully describe the orientation of the stellar rotational axis. 
From the stellar and planetary orbital geometry and the stellar deformation, we infer the local stellar temperature behind the planetary disc. 
We then convert this temperature into a fractional light loss (or gain) compared to the nominal
transit model, assuming black-body radiation and integrating over the CHEOPS' response function. 

We fit these angles ($I_\ast$, $\Omega_\ast$) together with the transit shape parameters, $R_{P}/R_{\ast}$, $b$, $T_0$, the relative 
semi-major axis, $a/R_{\ast}$, and the linear combinations of the quadratic limb-darkening coefficients, $u_+ =  u_a + u_b $ and $u_- =  u_a - u_b $. 
We assume a circular orbit and fix the period to that measured by \citet{Anderson18}.
The roll-angle-dependent flux variation is accounted for through a baseline model in form of a fourth-order Fourier series for each 
light curve and we allow for a constant normalisation offset.
As described in \citet{Csizmadia20}, we first explored a wide parameter space using a series of genetic algorithm and simulated annealing chains, 
before using the best solution found as a starting point for five independent MCMC chains of $10^6$ steps each. 
The convergence was checked through the \citet{Gelman92} statistic. 

We find a projected stellar obliquity of $\lambda = 86.4^{+2.9}_{-4.4}$\,deg. 
The true obliquity $\Psi$ - the angle between the stellar rotational axis and the orbital angular momentum vector - can be calculated via
\begin{equation}
  \cos \Psi = \cos I_{\ast} \cos i + \sin I_{\ast} \sin i \cos \lambda  \, ,
\end{equation}
and we find a value of $\Psi = 85.4\pm4.3$ deg. Here, $I_{\ast}$ and $i$ are the inclinations of the stellar rotational 
axis and the planetary orbit, respectively.
The projected and true obliquity values found here are in good agreement with the findings of \citet{Anderson18}, who reported values of $\lambda=89.3\pm1.4$~deg and $\Psi = 90\pm5.8$~deg based on spectroscopic measurements.

\begin{figure}
\centering
\includegraphics[width=\linewidth]{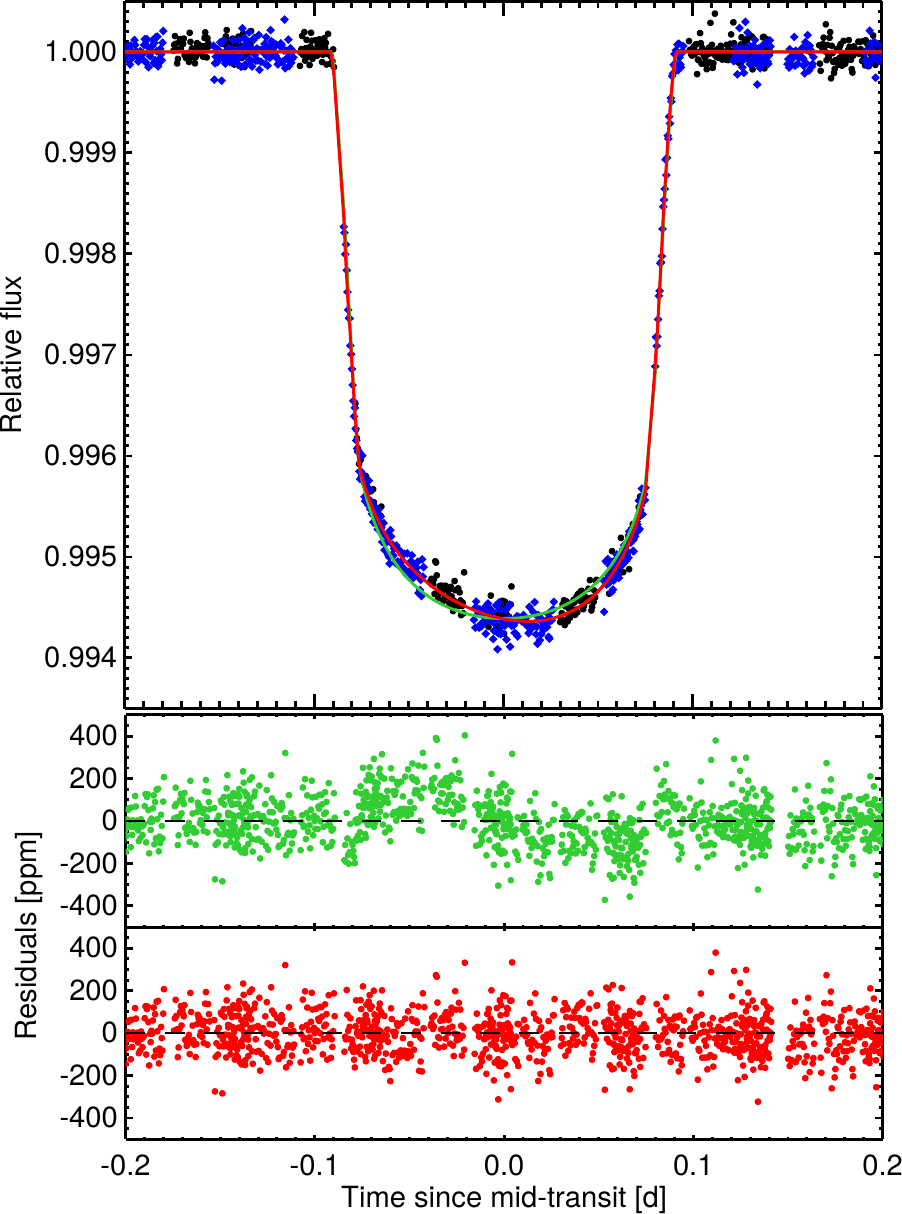}
\caption{\label{fig:lctransbin}Top: Corrected and phase-folded transit light curve of WASP-189\,b. Data from the 15 June 2020 are shown as black circles and data from 18 June 2020 are shown as blue diamonds.
The red and green curves illustrate the best-fit models, including and excluding gravity darkening, respectively. Bottom: Data residuals related to each of the models. Green 
points in the upper panel refer to residuals in the model without gravity darkening and red points in the lower panel refer to that with gravity darkening.}
\end{figure}

We list all inferred and derived parameters in Table \ref{tab:par}. The full list of baseline function coefficients for transits and occultations is given in Appendix \ref{app:coeff}. 
The individual and phase-folded transit light curves, together with the best-fit model, are shown in Figures \ref{fig:lctrans} and \ref{fig:lctransbin}, respectively. 
For the sake of comparison, we also show a model fit obtained by assuming a spherical star without gravity darkening in Figure~\ref{fig:lctransbin} (green curve).
It is evident from the residuals that the full model provides an improved fit for the asymmetric transit shape.

\begin{table}
\centering                        
\caption{\label{tab:par} Summary of stellar, input, and derived parameters of the WASP-189 system.
$^a$ fixed ; 
$^b$ $T_{\mathit{eq}}=\teff \sqrt{R_\ast / a} \,\, (f(1-A_{\rm B}))^{1/4}$, 
assuming immediate re-radiation ($f=2/3$) and zero albedo ($A_{\rm B}=0$) ;
$^c$ assuming black body stellar and planetary SEDs ;
$^d$ assuming a PHOENIX stellar model spectrum, $A_g = 0$, and inefficient energy circulation ($\epsilon = 0$).
}
\begin{tabular}{lc}       
\hline\hline 
 \multicolumn{2}{l}{Fitted parameters} \T  \\
\hline
Mid-transit time ($T_0$)                                     & $8926.5416960^{+0.000065}_{-0.000064}$  \T\\  
\quad [BJD$_{\mathit{TT}}$ -2450000]                         &         \\ 
Impact parameter ($b$)                                       & $0.478^{+0.009}_{-0.012}$     \T\\  
Scaled semi-major axis ($a/R_\ast$)                          & $4.60^{+0.031}_{-0.025}$      \T\\  
Eclipse duration ($T_{\mathit{14}}$) [h]                 & $4.3336^{+0.0054}_{-0.0058}$  \T\\  
Occultation depth ($dF_\mathit{occ}$) [ppm]                  & $87.9 \pm 4.3  $              \\ 
Radius ratio ($R_p/R_\ast$)      \T                          & $0.07045^{+0.00013}_{-0.00015}$ \T\\   
$u_{+} = u_a + u_b $                                         & $0.550^{+0.016}_{-0.017}$ \T\\  
$u_{-} = u_a - u_b$                                          & $0.440^{+0.066}_{-0.065}$ \T\\   
Stellar inclination $I_{\ast}$ [deg]                         & $75.5^{+3.1}_{-2.2}$      \T\\   
Projected orbital obliquity $\lambda$  [deg]                 & $86.4^{+2.9}_{-4.4}$      \T\\   
\hline
\multicolumn{2}{l}{Additional input parameters\T}\\
\hline
RV amplitude ($K$) [$\mathrm{kms^{-1}}$]                     & $0.182 \pm 0.013$ \T \\
Planetary period$^{a}$ ($P$) [d]                             & 2.7240330         \\
Eccentricity$^{a}$ ($e$)                                     & 0                 \\
\hline
\multicolumn{2}{l}{Stellar parameters\T}\\
\hline
Stellar Mass ($M_\ast$) [\Msolar]                            & $2.030 \pm 0.066$ \T \\
Stellar Radius ($R_\ast$) [\Rsolar]                          & $2.36 \pm 0.030$  \\
Stellar eff. temperature ($\teff$) [K]              & $8000 \pm 80$     \\
Stellar surface gravity log\,$g$ [\logg]            & $3.9 \pm 0.2$     \\
Projected rotational velocity                       & $93.1\pm1.7$      \\
\quad $\nu$sin$I_\ast$ [km\,s$^{-1}$]               &                   \\
Microturbulent velocity                             & $2.7 \pm 0.3$     \\
\quad $\nu_{\rm mic}$ [km\,s$^{-1}$]                &                   \\
Iron abundance [Fe/H]                               & $+0.29\pm0.13$    \\
System age [Myr]                                    & $730 \pm 130$     \\
\hline
\multicolumn{2}{l}{Derived parameters}\T\\
\hline
Plan. radius ($R_P$) [$\Rjup$]                               & $1.619\pm0.021   $ \T\\  
Plan. mass ($M_P$) [$\Mjup$]                                 & $1.99_{-0.14}^{+0.16}    $  \T\\ 
Plan. mean density ($\rho_P$) [\rhojup]                      & $0.469^{+0.058}_{-0.0275}$  \T\\  
Plan. surface gravity ($g_P$) [$\mathrm{ms^{-2}}$]           & $18.8^{+2.1}_{-1.8}      $  \T\\  
Orbital semi-major axis ($a$) [au]                           & $0.05053\pm0.00098       $  \\  
Orbital inclination ($i$) [deg]                              & $84.03 \pm 0.14          $  \\  
True orbital obliquity $\Psi$ [deg]                          & $85.4\pm4.3$                \\   
Dayside equilibrium temp.$^b$ ($T_{\mathit{eq}}$) [K]        & $3353_{-34}^{+27}$          \T\\
Brightness temp.$^c$ ($T_{\mathit{b}}$) [K]                  & $3348_{-35}^{+26}$          \T\\
Dayside temp.$^d$ ($T_{\mathit{day}}$) [K]                   & $3435 \pm 27$               \\
\hline
\hline
\end{tabular}
\end{table}

\section{Results} 
\label{sec:res}

\subsection{Revised planetary and system parameters}
The new,  high-precision CHEOPS observations allow us to substantially revise the planetary parameters, and the gravity-darkened nature of the stellar photosphere 
allows us to derive an independent measurement of the projected angle between the stellar spin and the planetary orbital axes. 

The remarkable difference of our results compared to those of \citet{Anderson18} is that we find a $\sim$25\% 
deeper transit, which is inconsistent with their published value at the level of 4.5$\sigma$. Paired with updated stellar parameters, this corresponds to a $\sim$15\% larger planetary radius 
(inconsistent at $2.9\sigma$) and, hence, a smaller planetary mean density.
We attribute this discrepancy to the difficulties in obtaining high-precision photometry 
for bright stars from the ground given that the quality of ground-based data for bright stars is limited by the paucity
of bright nearby reference stars. The photometric follow-up presented in \citet{Anderson18} is, furthermore, limited to partial transits, which often suffer from 
imprecisely determined photometric trends that can bias the observed transit depth. In contrast, neither the time trends related to stellar variability nor the roll-angle-dependent, in-orbit
variations in CHEOPS data exhibit amplitudes that are large enough to create a transit depth offset of the observed magnitude. Furthermore, as described in Section \ref{sec:obsCH}, 
the CHEOPS DRS has been validated on well-known planetary transits. 

From our gravity darkening analysis, we confirm a strongly misaligned orbit. While the analysis of the Rossiter-McLaughlin effect 
by \citet{Anderson18} yields $\lambda = 89.3\pm1.4$\,deg, our purely photometric 
analysis results in $\lambda = 86.4^{+2.9}_{-4.4}$\,deg. Assuming that the star rotates more slowly 
than its break-up velocity, \citet{Anderson18} find a true obliquity of $\Psi=90.0^\circ \pm 5.8^\circ$. 
Our photometric analysis is able to provide an assumption-free value of $\Psi = 85.4^\circ\pm4.3$.

\subsection{CHEOPS occultation measurement}

Based on a joint analysis of the four CHEOPS occultations, we determined the occultation depth of WASP-189\,b in the CHEOPS passband to be $87.9\pm4.3$\,ppm.
The precision of this measurement exceeds that of previous measurements obtained with \emph{CoRoT} \citep{Parviainen13}, 
and TESS \citep[see][and references therein]{Wong20}, and is comparable in precision with the occultation depth measurements of hot Jupiters inferred from several quarters worth of 
Kepler data \citep[e.g.][]{Angerhausen15,Esteves15,Morris13}.

The individual, unbinned, occultation light curves, which have a cadence of 33.4\,s, have a residual RMS between 86 and 92\,ppm. 
When applying binning into 10-minute and 1-hour intervals, we reach RMS values between 34 and 47, and 10 and 17\,ppm, respectively. 
The phase-folded and binned residuals show an RMS of 23 and 5.7\,ppm for 10-minute and 1-hour time bins, respectively. 
These values underline the excellent performance of CHEOPS.

Motivated by the high level of precision reached here, we also carried out independent analyses of each occultation to probe for any potential variation 
in the measured occultation depth. The occultation is detected at high significance in each individual light curve and the measurements are consistent at
1-$\sigma$ level. Thus, we find no significant sign of variability (see Table \ref{tab:sep}) in the dayside flux from WASP-189\,b over the 19-day time span of our observations.
At the same time, this illustrates that the value derived from a joint fit is not biased by any individual light curve.

\begin{table}
\centering                        
\caption{\label{tab:sep}Occultation depths inferred from analyses of individual visits.}
\begin{tabular}{lcccc}       
\hline\hline 
Date (all 2020) & 19 Mar& 27 Mar & 30 Mar & 7 Apr \T\\
\hline 
$\mathit{dF_{occ}}$ [ppm] & $88.6_{-11}^{+8.5}$ & $83.5_{-8.5}^{+11.4}$ & $94.1_{-9.6}^{+9.9}$ & $89.3_{-6.9}^{+6.5}$\T \\
\hline\hline
\end{tabular}
\end{table}

\begin{figure*}[!h]
\centering
\includegraphics[width=0.49\linewidth]{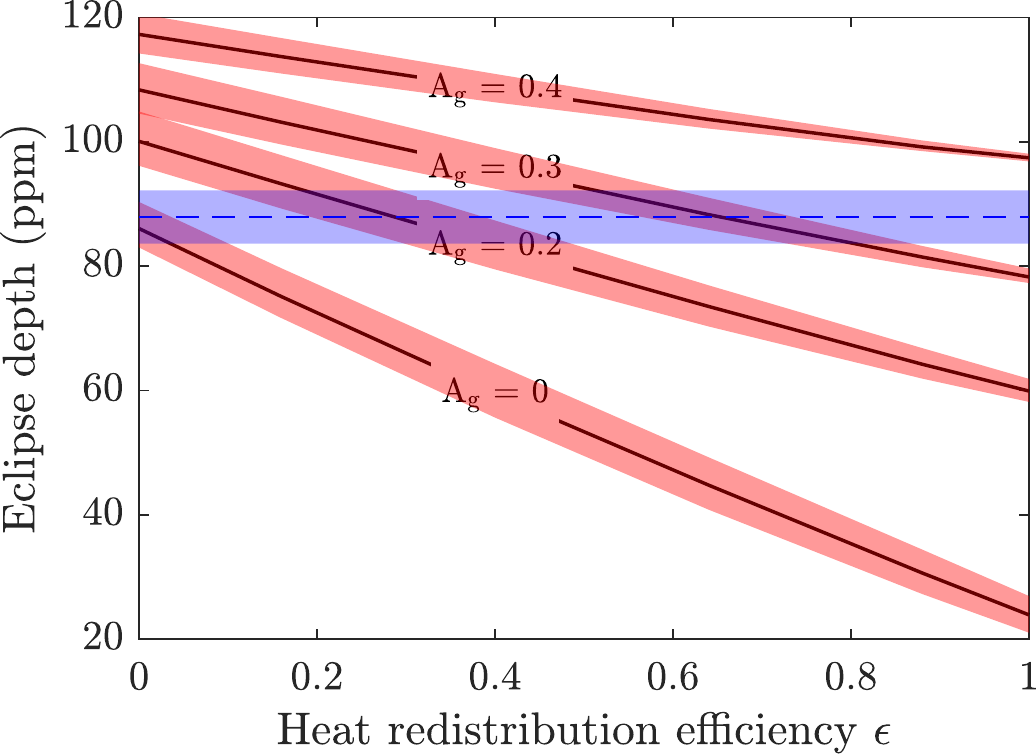}
\includegraphics[width=0.49\linewidth]{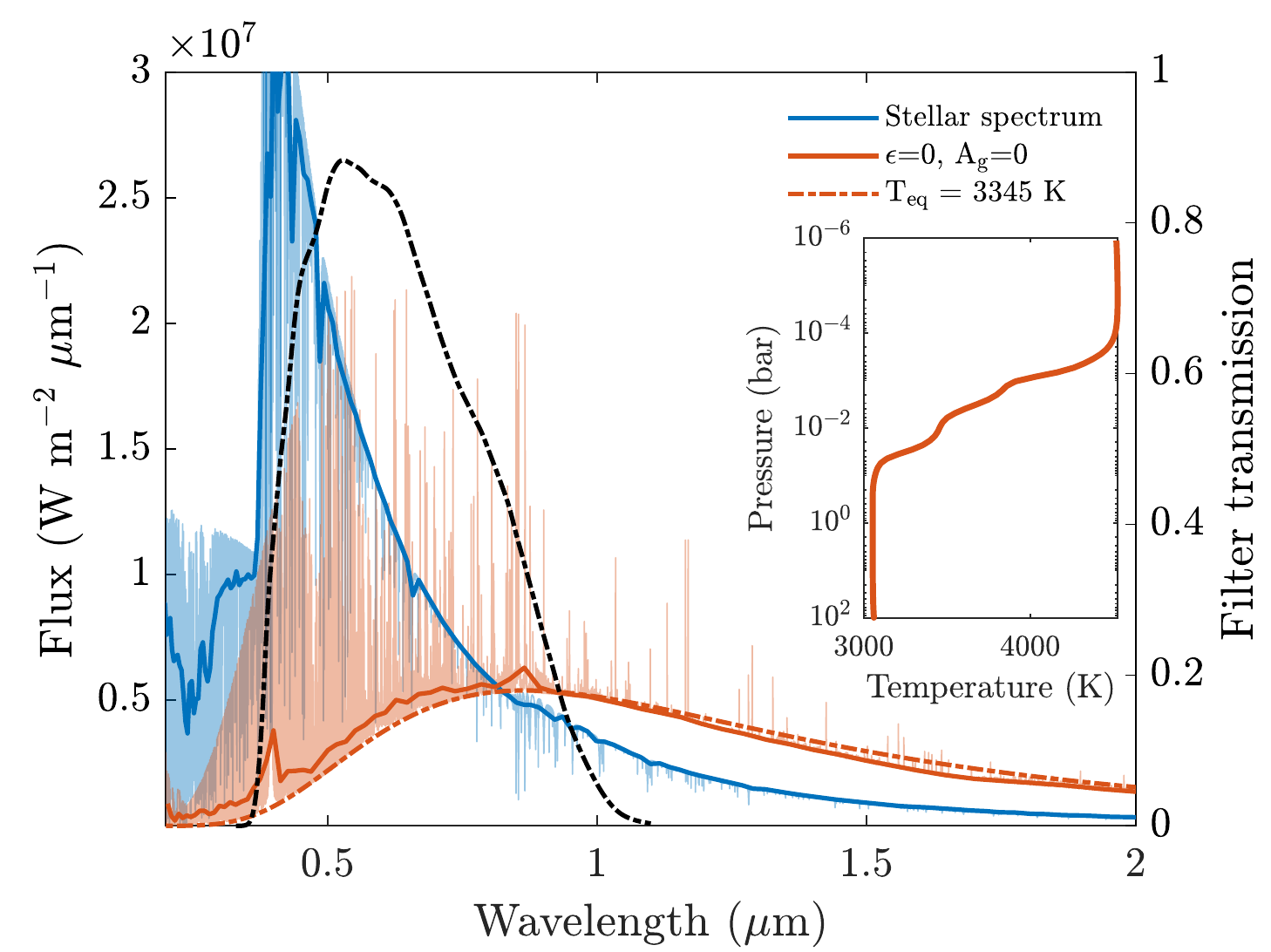}
\caption{Left panel: Calculated (curves) versus measured (shaded horizontal region) occultation depths as functions of the heat redistribution efficiency. 
Different curves with assumed values of $A_g$ are shown. As a sensitivity test, the shaded region associated with each curve corresponds to a 
variation in metallicity within a range of $[\mathrm M/ \mathrm H] = 0.2 \pm 0.3$.  
Right panel: Theoretical spectral energy distribution, at a low and high resolution of the star (blue curve) with the CHEOPS bandpass (black dot-dashed curve) overlaid.
The model for WASP-189\,b (with $A_g=\epsilon=0$) is overlaid in orange, with the corresponding temperature-pressure profile shown in the inset.
For comparison, a blackbody curve with a temperature of 3345\,K is also overlaid (orange dashed line). }
\label{fig:model}
\end{figure*} 

\subsection{The atmosphere of WASP-189\,b}
\subsubsection{Model description}
To interpret the occultation depth, the radiative transfer code \texttt{HELIOS} was used to calculate the spectral energy distribution (SED) 
of the dayside atmosphere of WASP-189\,b. \texttt{HELIOS} solves for the thermal structure self-consistently \citep{Malik17,Malik19}. 
The model atmosphere is assumed to be cloud-free and in chemical equilibrium. We varied the planet's atmospheric metallicity 
within $[\mathrm M/ \mathrm H] = 0.2 \pm 0.3$, based on the stellar abundances. Sources of opacity include: 
spectral lines of atoms and ions of metals (Ca, Ca$^+$, Fe, Fe$^+$, Ti, Ti$^+$, Na, K; \citealt{Kurucz95}), 
which are predicted theoretically \citep[e.g.][]{Kitzmann18} and observed at a high resolution in other ultra-hot Jupiters \citep[e.g.][]{Hoeijmakers19}; 
spectral lines of H$_2$O, CO, CH$_4$, 
VO and TiO \citep{Barber2006_10.1111/j.1365-2966.2006.10184.x, Yurchenko2014MNRAS.440.1649Y, Rothman2010JQSRT.111.2139R, McKemmish2016MNRAS.463..771M, McKemmish2019MNRAS.488.2836M}; 
continuum absorption from the hydrogen anion (H$^-$; \citealt{John88}); H$_2$-H$_2$, H$_2$-He and H-He collision-induced absorption \citep{Karman2019Icar..328..160K}. 
It is worth noting that \texttt{HELIOS} includes albedo contributions from Rayleigh scattering due to molecules. 
As illustrated in Figure \ref{fig:model}, our models predict that WASP-189\,b possesses a thermal inversion, as inferred recently by \citet{Yan20} from high-resolution spectroscopic observations.
We report the planetary dayside temperature in Table \ref{tab:par}, next to the brightness temperature computed 
under the assumption of black-body emission for star and planet. As described in 
Appendix \ref{app:Tb}, these are discrepant because the assumption of black-body emission is flawed due to the proximity of the CHEOPS band to the Balmer jump.

The measured occultation depth can be explained by a combination of thermal emission and a weakly-reflective atmosphere (i.e. geometric albedo $A_g \sim \left[0.1-0.3\right]$) for most values
of the heat redistribution efficiency ($\epsilon$, see below). We note that thermal emission alone ($A_g=0$) 
may account for the measured occultation depth if zero heat redistribution is assumed ($\epsilon = 0$).

\subsubsection{Scattering by clouds/hazes}
Since the heat redistribution efficiency ($\epsilon$) is unknown, a broader interpretation of the measured occultation depth may be obtained by assuming that scatterers of unknown origin and composition which are associated with clouds or hazes are present in the model atmosphere. They are parameterised by a single 
value of the geometric albedo ($A_g$). The occultation depth has contributions from reflected light and thermal emission,
namely,\begin{equation}
dF_{occ} = A_g \left( \frac{R_p}{a} \right)^2 + \frac{\int {\cal F} F_p ~d\lambda}{\int {\cal F} F_*  ~d\lambda} \left( \frac{R_p}{R_*} \right)^2.
\end{equation}
The CHEOPS bandpass (${\cal F}$), the SED of the star ($F_*$, as computed in Section \ref{sec:star}) and an example of the SED of WASP-189\,b ($F_p$) are shown in Figure \ref{fig:model}. 
As an input to \texttt{HELIOS}, the top-of-the-atmosphere (TOA) flux impinging upon WASP-189\,b is
\begin{equation}
F_{\rm TOA} = F_* \left( \frac{R_*}{a} \right)^2 \left( 1 - A_{\rm B} \right) \left( \frac{2}{3} - \frac{5\epsilon}{12} \right),
\end{equation}
where the heat redistribution efficiency ($0 \le \epsilon 
\le 1$) follows the parametrisation of \citet{Cowan11}. It is related to the commonly used redistribution factor of $1/4\le f \le 2/3$  \citep{Seager05} via $\epsilon = 8/5 - 12f/5$. 
To relate the geometric and Bond ($A_{\rm B}$) albedos, isotropic scattering is assumed such that $A_g = 2A_{\rm B}/3$.

Figure \ref{fig:model} shows that $A_g \sim 0.1$ models are easily consistent with the measured occultation depth if $\epsilon \sim 0.1$, which is, in turn, consistent with the values of geometric albedos measured for cooler hot Jupiters \citep{Heng13}.

\section{Conclusions and outlook}
\label{sec:con} 

In this paper, we present CHEOPS observations of the hot Jupiter WASP-189\,b, capturing both the transit and the occultation of the highly irradiated planet. 
We robustly detect the occultation in individual epochs and measure a depth of $87.9 \pm 4.3$~ppm 
when combining four occultation light curves. Our measurement can be reproduced by atmospheric models with comparatively low albedo and heat redistribution efficiency. 
From two transit light curves, we derive updated planetary parameters and find a $\sim$15\% larger planetary radius.
The transits clearly show an asymmetric shape due to gravity darkening of the stellar host, and we use this effect to measure the planetary spin-orbit angle, finding
a clearly misaligned orbit with a projected obliquity of $\lambda =  86.4^{+2.9}_{-4.4}$\,deg and a true obliquity of $\Psi = 85.4 \pm4.3$\,deg.

These observations showcase the capability of CHEOPS to detect shallow signals with an extremely high level of precision, thereby illustrating the potential of 
future studies of exoplanet atmospheres with CHEOPS. 
These will include (geometric) albedo measurements for cool planets, which have negligible contribution of thermal emission in the optical, as well as for 
planets, which have a dayside emission spectrum that is well-known from infra-red observations. For the most favourable objects, CHEOPS will conduct phase curve observations, 
revealing the longitudinal cloud distribution in the planets' atmosphere.
Thanks to its flexible pointing and observing schedule, CHEOPS can point to exoplanets across large 
areas of the sky, targeting the most rewarding objects. These practical aspects make CHEOPS an ideal facility for collecting a large sample of optical-light exoplanet 
occultations and phase curves.

\begin{acknowledgements}
CHEOPS is an ESA mission in partnership with Switzerland with important contributions to the payload and the ground segment from Austria, Belgium, France, 
Germany, Hungary, Italy, Portugal, Spain, Sweden, and the United Kingdom.
The Swiss participation to CHEOPS has been supported by the Swiss Space Office (SSO) in the framework of the Prodex Programme and the 
Activit\'es Nationales Compl\'ementaires (ANC), the Universities of Bern and Geneva as well as well as of the NCCR PlanetS and the Swiss National Science Foundation.
MLE acknowledges support from the Austrian Research Promotion Agency (FFG) under project 859724 ``GRAPPA''.
Sz. Cs. thanks DFG Research Unit 2440: ’Matter Under Planetary Interior Conditions: High Pressure, Planetary, and Plasma Physics’ for support. Sz. Cs. acknowledges support by DFG grants RA 714/14-1 within the DFG Schwerpunkt SPP 1992: 'Exploring the Diversity of Extrasolar Planets'.
ADE and DEH acknowledge support from the European Research Council (ERC) under the European Union’s Horizon 2020 research and innovation programme 
(project {\sc Four Aces}; grant agreement No 724427).
MJH acknowledges the support of the Swiss National Fund under grant 200020\_172746.
The Spanish scientific participation in CHEOPS has been supported by the Spanish Ministry of Science and Innovation and the 
European Regional Development Fund through grants ESP2016-80435-C2-1-R, ESP2016-80435-C2-2-R, ESP2017-87676-C5-1-R, 
PGC2018-098153-B-C31, PGC2018-098153-B-C33, and MDM-2017-0737 Unidad de Excelencia Mar\'{\i}a de Maeztu--Centro de Astrobiolog\'{\i}a (INTA-CSIC), 
as well as by the Generalitat de Catalunya/CERCA programme. The MOC activities have been supported by the ESA contract No. 4000124370.
This work was supported by FCT - Funda\c{c}\~ao para a Ci\^encia e a Tecnologia through national funds and by FEDER through 
COMPETE2020 - Programa Operacional Competitividade e Internacionaliza\c{c}\~ao by these grants: 
UID/FIS/04434/2019; UIDB/04434/2020; UIDP/04434/2020; PTDC/FIS-AST/32113/2017 \& POCI-01-0145-FEDER-032113; 
PTDC/FIS-AST/28953/2017 \& POCI-01-0145-FEDER-028953; PTDC/FIS-AST/28987/2017 \& POCI-01-0145-FEDER-028987. 
S.C.C.B. and S.G.S. acknowledge support from FCT through FCT contracts nr. IF/01312/2014/CP1215/CT0004, IF/00028/2014/CP1215/CT0002. 
O.D.S.D. is supported in the form of work contract (DL 57/2016/CP1364/CT0004) funded by national funds through Funda\c{c}\~ao para a Ci\^encia e Tecnologia (FCT).
The Belgian participation to CHEOPS has been supported by the Belgian Federal Science Policy Office (BELSPO) in the framework of the PRODEX Program, 
and by the University of Liege through an ARC grant for Concerted Research Actions financed by the Wallonia-Brussels Federation. MG is F.R.S.-FNRS Senior Research Associate.
S.S. has received funding from the European Research Council (ERC) under the European Union's Horizon 2020 research and innovation programme (grant agreement No 833925, project STAREX).
GyS acknowledges funding from the Hungarian National Research, Development and Innovation Office (NKFIH) grant GINOP-2.3.2-15-2016-00003 and K-119517.
For Italy, CHEOPS activities have been supported by the Italian Space Agency, under the programs: ASI-INAF n. 2013-016-R.0 and ASI-INAF n. 2019-29-HH.0.
The team at LAM acknowledges CNES funding for the development of the CHEOPS DRP, including grants 124378 for O.D. and 837319 for S.H.
XB, SC, DG, MF and JL acknowledge their role as an ESA-appointed CHEOPS science team members.
\end{acknowledgements}

\bibliographystyle{aa}
\bibliography{bbl}

\begin{appendix}

\section{Stellar abundances}
\label{app:abundances}
\begin{figure}[h!]
\centering
\includegraphics[width=9cm]{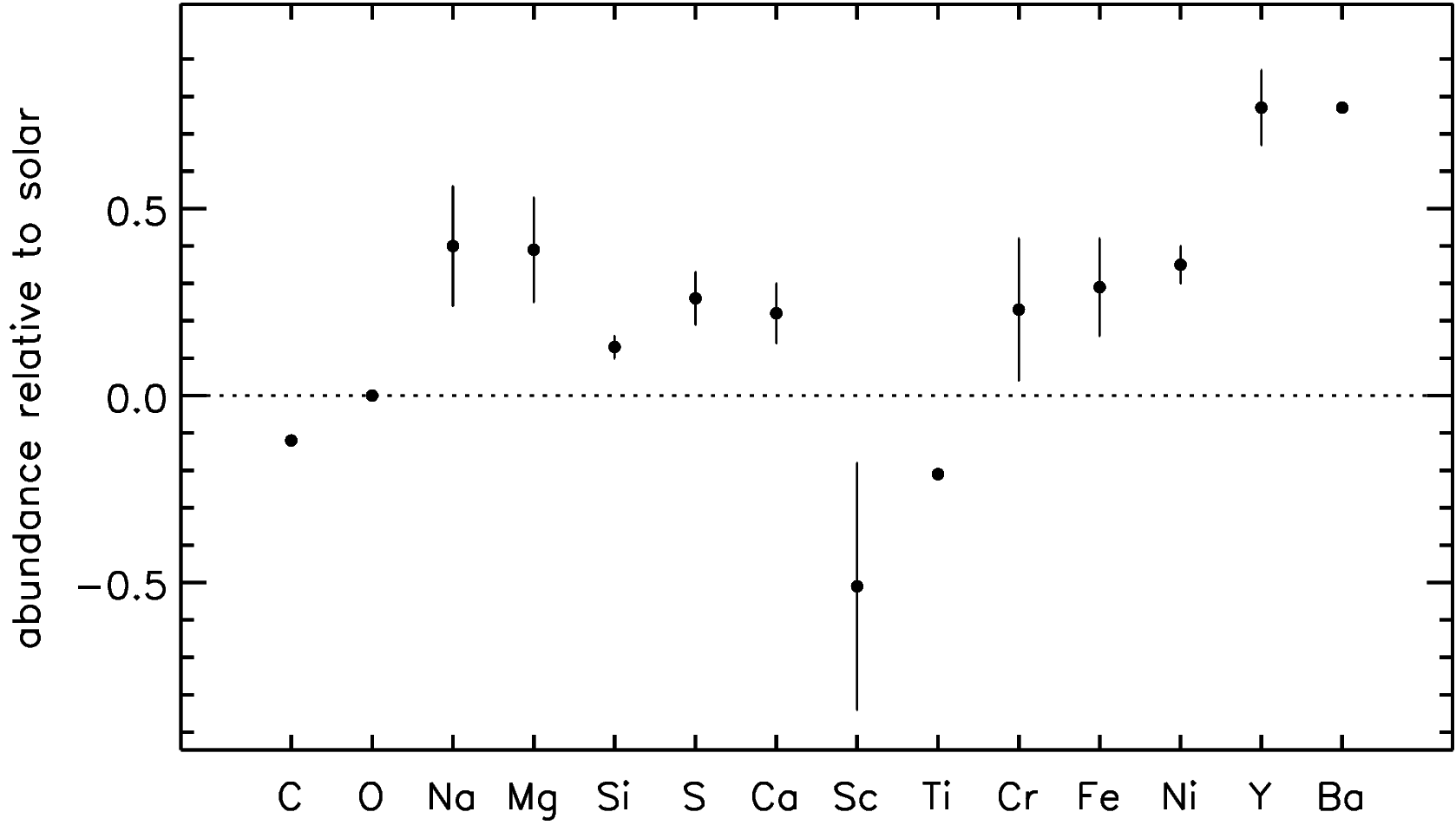}
\caption{WASP-189 abundance pattern. The abundances are relative to solar \citep{Asplund09}. 
The uncertainties are the standard deviation from the average abundance, therefore the abundances obtained from only one line 
(C, O, Ti, Ba) are shown without uncertainties. 
}
\label{fig:abundances}
\end{figure}

The stellar abundance pattern is derived using the methods described in Section \ref{sec:star} and displayed in Figure \ref{fig:abundances}.

\section{Photometric baseline model parameters}
\label{app:coeff}
In Table \ref{tab:det}, we report the inferred parameters and uncertainties for the baseline model parameters of each individual light curve.

\begin{table}
\centering                        
\caption{\label{tab:det}
Coefficients found for the photometric baselines models fitted jointly with the physical light curve model.
For the occultations, $A_i$ refer to the coefficients of second-order polynomials in time, with
$A_0$ denoting the normalisation constant. $D_1$ is the coefficient of a linear trend with contamination.
For the transits, $c$ stand for the cosine, and $s$ for the sine terms of the Fourier-series.}
\begin{tabular}{lcc}       
\hline
\hline
\multicolumn{3}{c}{Occultations \T} \\
\hline
Date & 19 Mar 2020 & 27 Mar 2020 \T \\
\hline
$A_0$ & $1.0000805_{-0.0000093}^{+0.0000086}$ & $1.000024_{-0.0000096}^{+0.000010} $ \\
$A_1$ & $-0.001221_{-0.000089}^{+0.000080} $  & $-0.001130_{-0.000101}^{0.000082} $  \\  
$A_2$ & $0.00162_{-0.00016}^{+0.00015} $      & $ 0.00184_{-0.00015}^{+0.00020} $    \\
$D_1$ & $3.20_{-0.16}^{+0.17} $               & $ 2.22_{-0.18}^{+0.15}$      \\
\hline
Date & 30 Mar 2020 & 07 Apr 2020 \T \\
\hline
$A_0$ & $1.000033_{-0.000011}^{+0.0000090} $ & $ 1.0000018_{-0.0000058}^{+0.0000066} $ \T \\
$A_1$ & $-0.000090_{-0.000090}^{+0.000086} $ & $ 0.000189 \pm 0.000019 $ \\
$A_2$ & $0.00028_{-0.00015}^{0.00016} $ & 0 \\
$D_1$ & $ 2.29 \pm0.16 $ & $1.95_{-0.16}^{+0.14}$ \\
\hline
\hline
\multicolumn{3}{c}{Transits \T} \\
\hline
Date & 15 Jun 2020 & 18 Jun 2020  \T\\
\hline
Flux shift [ppm] & \multicolumn{2}{c}{$-61\pm7$} \T  \\
$c_1$ [ppm]  & $-76\pm14$   & $-30\pm14$  \\
$c_2$ [ppm]  & $+74\pm11$   & $-5\pm10$  \\
$c_3$ [ppm]  & $+78\pm16$   & $+23\pm17$ \\
$c_4$ [ppm]  & $+18\pm9$   & $+6\pm9$  \\
$s_1$ [ppm]  & $-40\pm14$  & $+90\pm11$  \\
$s_2$ [ppm]  & $-74\pm19$  & $+17\pm21$  \\
$s_3$ [ppm]  & $+3\pm11$  & $-13\pm11$ \\
$s_4$ [ppm]  & $+26\pm9$  & $-16\pm10$ \\
\hline
\end{tabular}
\end{table}

\section{Planetary brightness temperature}
\label{app:Tb}
\begin{figure}[h!]
\centering
\includegraphics[width=\linewidth]{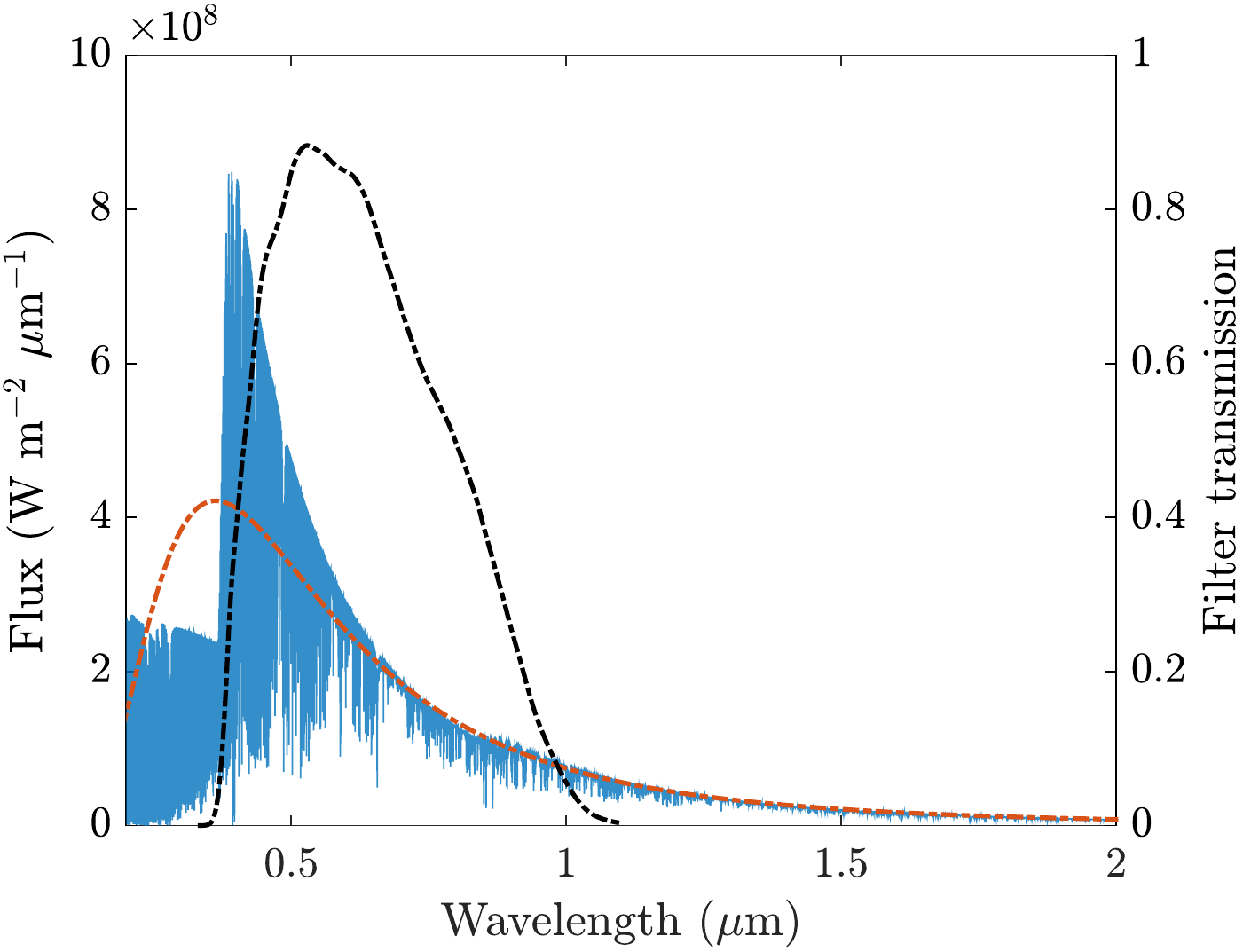}
\caption{\label{fig:SEDs}Comparison of a PHOENIX \citep{Husser13} stellar spectrum for a star with parameters corresponding to WASP-189 (blue), a 8000\,K black-body (orange), 
and the CHEOPS passband (black).}
\end{figure}

We remark that, unlike the case of long-wavelength measurements, approximating the stellar emission by a black-body SED leads to an under-estimation of the stellar flux 
in the CHEOPS passband and, thus, it under-estimates the planetary dayside temperature. This is illustrated in Figure \ref{fig:SEDs}, which shows a model stellar spectrum compared to 
emission from a 8000\,K black-body. The difference is attributed to the proximity of the CHEOPS band to the Balmer jump. For the case of WASP-189\,b, we find a brightness 
temperature of $3348_{-35}^{+26}$ when using the black-body approximation, but a higher value of $3435 \pm 27$\,K when using a stellar model spectrum.

\end{appendix}

\end{document}